\documentclass{article}
\def\upcirc#1{\vbox{\ialign{##cr
$\circ\!$cr \noalign{\kern-0.1
pt\nointerlineskip}
$\hfil\displaystyle{#1}\hfil$cr}}}

\def\thorn{\hbox{\rm I}\kern-0.32em\raise0.35ex\hbox{\it o}}
\def\edth{\hbox{$\partial$\kern-0.25em\raise0.6ex\hbox{\rm\char'40}}}

\def\i{\iota}

\def\thornp{{\thorn^\prime}}
\def\edthp{{\edth^\prime}}

\def\pmb#1{\setbox0=\hbox{#1} \kern-.025em\copy0\kern-\wd0
\kern.05em\copy0\kern-\wd0
\kern-.025em\raise.0233em\box0 }
\def\bedth{{\pmb{\edth}}}
\def\bthorn{{\pmb \thorn}}
\def\thornp{{\thorn^\prime}}
\def\edthp{{\edth^\prime}}

\def\bthornp{{\bthorn^\prime}}
\def\bedthp{{\bedth^\prime}}

\def\I{\bf I}

\def\Ph{\nthorn}
\def\D{\nedth}
\def\pmb#1{\setbox0=\hbox{#1}%
  \kern-.025em\copy0\kern-\wd0
  \kern.05em\copy0\kern-\wd0
  \kern-0.025em\raise.0433em\box0 }
\def\half{{\scriptstyle {1 \over 2}}}
\def\third{{\scriptstyle {1 \over 3}}}
\def\quarter{{\scriptstyle {1 \over 4}}}
\def\o{{\pmb o}} %this should be omicron
\def\i{{\pmb{$\iota$}}}
\let\gt=\mapsto
\let\la=\lambda

\def\ob{{\bar\o}}
\def\ib{{\bar\i}}
\def\T{{\bf T}}
\def\bp{{\bf p}}
\def\bq{{\bf q}}
\def\0{\pmb 0} % should be bold 0
\def\1{\pmb 1} % should be bold 1
\def\2{\pmb 2} % should be bold 1
\def\3{\pmb 3} % should be bold 1
\def\bPhi{{\pmb{$\Phi$}}}
\def\bPsi{{\pmb{$\Psi$}}}
\def\>{\phantom{A}}
\def\sym{{\sum_{sym}}}
\def\thorn{\hbox{\rm I}\kern-0.32em\raise0.35ex\hbox{\it o}}
\def\edth{\hbox{$\partial$\kern-0.25em\raise0.6ex\hbox{\rm\char'40}}}

\def\nedth{{\pmb\edth}}
\def\nedthp{{\pmb\edth'}}
\def\thornp{\thorn'}
\def\nthorn{{\pmb\thorn}}
\def\nthornp{{\pmb\thorn'}}
\def\edthp{\edth'}

\def\cd{{\cal D}}
\def\boeta{{\pmb{$\eta$}}}

\def\etad{\boeta_{C_1\dots C_N C'_1\dots C'_{N'}}}
\def\p{{\bf p}}
\def\q{{\bf q}}
\def\K{{\bf K}}
\def\R{{\bf R}}
\def\S{{\bf S}}
\def\T{{\bf T}}
\def\I{{\bf I}}
\def\la{\lambda}
\def\lab{\bar\lambda}

\input{tcilatex}

\begin{document}

\title{Obtaining a class of Type N pure radiation metrics using invariant
operators.}
\author{S. Brian Edgar
\\ Department of Mathematics,
 \\ Link\"{o}pings Universitet\\ Link\"{o}ping,\\
Sweden S-581 83\\
 email: bredg@mai.liu.se
\and M.P.Machado Ramos
\\ Departamento de Matem\'{a}tica para a Ci\^{e}ncia e Tecnologia,
\\Azur\'{e}m 4810 Guimar\~{a}es,
\\ Universidade do Minho, \\ Portugal.\\
 email: mpr@mct.uminho.pt}
 \maketitle

\section*{Abstract.}

We develop further the integration procedure in the  generalised invariant formalism, and demonstrate its efficiency by obtaining a class of Petrov type N  pure radiation metrics without any explicit integration, and with comparatively little detailed calculations. The
method is similar to the one exploited by Edgar and Vickers when
deriving the general conformally flat pure radiation metric. A major addition to the technique is the introduction  of non-intrinsic elements in generalised invariant formalism, which can be exploited to keep calculations manageable.

{\bf PACS numbers:} \ 0420,\ 1127

\section{Introduction}

The NP formalism \cite{np} has been a very useful tool for
obtaining and analysing solutions of Einstein's equations. The tetrad
freedom inherent in this formalism has been both a boon and a hindrance;
often a skillful tetrad choice gives an immediate simplification which
sometimes leads directly to a satisfactory conclusion, but at other times
the original simplification proves to be cosmetic and an impass is
reached in the calculations. There is usually an obvious choice for the
direction of the first null tetrad vector; sometimes the direction of the
second null tetrad vector suggests itself; but there is usually no
geometric motivation to fix the spin and boost freedom. An additional
disadvantage is that the exploitation of the tetrad freedom, especially
that associated with null rotation and spin and boost after the first null
direction is fixed, can require long painstaking calculations, keeping
track as intertwined tetrad and coordinate freedoms are gradually used up
in successive steps; the possibility of errors is always present.

The GHP formalism \cite{ghp}, which generalises the NP formalism, has the spin and
boost freedom inbuilt (as spin and boost 'weight') which means that we do
not need to make an explicit choice of spin and boost gauge; so this is the
ideal formalism to use when investigating spaces which pick out two null directions in a
geometric manner.  Moreover, by rescaling all quantities in the GHP
formalism to zero-weighted quantities, together with only one complex
non-zero weighted quantity, we are able to seperate out the effect of the
spin and boost gauge and essentially ignore it; working  exclusively
with zero-weighted quantities gives natural coordinates and results which
are more accessible to  physical interpretation. In addition the messy
calculations often associated with fixing the spin and boost gauge  do
not occur.

However, investigations of those spacetimes where only the
direction of one null vector is picked out geometrically, often
encounter the complication associated with the freedom of a null
rotation. This freedom can cause analogous computational
complications in a GHP analysis as the spin and boost freedom case
in a NP analysis.

The generalised invariant formalism (GIF) \cite{maria1},
\cite{maria2} generalises the GHP formalism by essentially
building the  null rotation freedom  into the GIF, which means
that the formalism is built around one dyad spinor $\o_A$, and we
do not need to make an explicit choice for the null rotation
freedom of the second dyad spinor $\i_A$. But the price we pay is
that we have to move from a scalar to a spinor formalism, which at
first sight appears very complicated, and even unwieldy. However,
manipulating within the GIF is in practice (although not in
appearance) not much more complicated than the scalar
manipulations of the GHP formalism.

\medskip

Some years ago Held \cite{held1}, \cite{held2} outlined an
idealised integration method in an 'optimal situation' using the
GHP formalism.  He advocated manipulating the GHP equations until
they were reduced to a complete and involutive  set of tables
involving  first  derivative GHP operators; from these tables Held
argued it should be possible to read off the metric {\it without
having to perform  any explicit integrations}. Subsequently this
idea has been developed and applied in a number of ways, and  it
is now understood that the 'optimal situation'   is when the
manipulation succeeds in generating a  table for each of four real
zero-weighted scalars (which will become the coordinates) and  a
table for one complex non-zero-weighted scalar (which describes
the spin and boost gauge) \cite{edGHP}. An important technique in
this method is the repeated application of the commutator
equations; however, in addition, it is crucial to recognise the
theoretical importance of the commutator equations where important
information resides (as well as in the GHP Ricci and Bianchi
equations) and in order that all this information has been
extracted it is essential that the commutators should have been
applied explicitly to these five scalars \cite{edGHP}. These five
tables will not in general be completely involutive and we would
expect some additional scalars and their (possibly, partial)
tables to ensure a complete and involutive system.

Following the pioneering work of Held \cite{held1}, \cite{held2}, the most striking illustrations of the advantages and efficiency of this GHP formalism have been illustrated
in a re-investigation  of Petrov D vacuum metrics \cite{carm1}, \cite{carm2} where key constraint equations were obtained with a fraction of the calculations required in the NP formalism,
and  in the re-investigation of the
conformally flat pure radiation metrics \cite{edlud} where new solutions were obtained,
which had been overlooked in the complexities of an earlier NP
investigation \cite{wils}.

In a recent paper by Edgar and Vickers \cite{edgar} this GHP
integration method was generalised to the GIF \cite{maria2}.
Again, the method consists of manipulating all the  equations of
the formalism to construct a complete and involutive  set of
tables  involving  first  derivative GIF spinor operators. The
'optimal  situation' to be sought is to generate a  table for each
of four real zero-weighted scalars (which will become the
coordinates), a  table for one complex (non-trivially-)weighted
scalar (which describes the spin and boost gauge) and a table for
a second  spinor $\I_A$; such a spinor (which is    not parallel
to the first dyad spinor $\o_A$) should   emerge naturally from
the calculations, and can then be identified as the second dyad
spinor $\i_A$.  Again, an important element in this method is to
recognise that much information resides in the GIF commutator
equations (as well as in the GIF Ricci and Bianchi equations) and
in order that all this information has been extracted it is
essential that the commutators should have been  applied
explicitly to these five scalars as well as to the new spinor
${\bf I}_A$. Once these tables have been found, and the new spinor
identified with the second dyad spinor $\i_A$, the problem can be
reduced to a purely scalar one in the GHP formalism.

The advantage of the GIF  was illustrated by a further
investigation of the conformally flat pure radiation metrics \cite{edgar}. In the GHP
investigation a degree of guesswork and luck was involved in obtaining
that particular  choice of null rotation which fixed the direction of the
second null vector in such a manner that subsequent calculations became
manageable.  In the GIF  no explicit choice was needed, since the
formalism deals only with quantities which are invariant under such null
rotations. Moreover, the resulting form for the class of metrics was such that Killing vector properties were obvious
 and  the invariant
classification procedure could easily  be deduced; such conclusions could not be made from the
form of the metric obtained by the GHP analysis. The two metric forms
differ only slightly  in appearance, but
fundamentally different in the interpretation of some of the coordinates.

\medskip

An intriguing aspect of this GHP and GIF integration method is
that it is best suited to spaces without any Killing vectors. The
principle of the method is to try to generate the four coordinates
directly from the {\it intrinsic} elements of  the formalism, and
in spaces with no Killing vectors the coordinates will be
generated in this manner. When Killing vectors are present there
are not enough intrinsic elements of the formalism to generate
four intrinsic coordinates, and so new {\it non-intrinsic}
elements have to be introduced.

 The conformally flat pure radiation metrics \cite{edlud}, \cite{edgar} provided an 'optimal situation' and gave a
 comparatively simple demonstration because we were able to generate all four cordinates
intrinsically from the elements of the formalism {\it in the
generic case} \cite{edgar}. However, there was a subclass of these
spaces for which we could only generate three coordinates
intrinsically (corresponding to the existence of a Killing
vector). We were able to identify these spaces by modifying the
technique and using the structure of the commutators to motivate a
coordinate choice from outside the intrinsic elements of the GIF;
we then explicitly confirmed that this choice was compatible with
all the other equations.

This technique of introducing a new non-intrinsic element into the
computations --- motivated by the property that it automatically
satisfies the  commutator equations and implies no new constraints
--- is essential when dealing with spaces in which Killing vectors are
present; however, this technique can also be used to try and
obtain simpler tables when the direct method generates
complicated ones.

So far, the conformally flat pure radiation metrics are  the only
class of metrics which have been determined explicitly by this GIF
procedure \cite{edgar}. The GIF formalism was initially designed
to be used to investigate the invariant classification of
spacetimes; in particular of those metrics where the second null
tetrad vector has no simple geometric links, e.g. Petrov type N
vacuum spaces \cite{maria3}. However, it is clear that the GIF can
also be used for obtaining solutions of Einstein's equations for
metrics with these properties, but before attacking very difficult
problems like the whole class of vacuum Type N twisting metrics,
it is necessary to build up more experience of using the GIF.

\medskip
Therefore, in an effort to better understand how the GIF
integration procedure works in practice we shall investigate a
class of Petrov type N pure radiation metrics. These metrics are
close enough to their conformally flat counterparts in
\cite{edgar} for us to get some hints of  how to tackle them in
the GIF, but sufficiently different to present some new
challenges; in particular we expect a richer Killing vector
structure, and hence a less direct method. Unlike in \cite{edgar}
where we followed a direct method, and only brought in one
coordinate candidate from outside the formalism at the last step,
in this approach we will introduce a new element from the
beginning. These metrics are well known, especially in their
familiar Kundt form \cite{kramer}. So the purpose of this paper is
not  to find new solutions, but rather to rederive a known class
by developing new techniques within the GIF procedure; in
particular to demonstrate how introducing non-intrinsic elements
can keep calculations simple and manageable.  We are further
motivated to demonstrate the power of the GIF approach for these
spaces because attempts to make an  investigation of these spaces
simply and efficiently in the GHP formalism have been unsuccessful
due to the problem of the null rotational freedom of the second
tetrad vector;  instead a very detailed and involved calculation
has been needed in NP formalism and in a hybid approach combining
GHP with NP ideas \cite{ludwig}.

 \medskip

 In the next section we summarise the GIF  \cite{maria2}, with special reference to differential operators.
 In Section 3 we specialise the GIF  to the class of spaces under discussion, and in Section 4 we carry out the
 GIF integration procedure for these spaces.  Section 5 gives a summary.

\section{The Differential Operators}

In the GIF  the role of the spin coefficients
$\kappa$, $\sigma$, $\rho$ and $\tau$ is taken up by spinor quantities
$\K$, $\S$, $\R$ and $\T$ given by
\begin{eqnarray}
\K&=&\kappa \nonumber\\
\S_{A^\prime}&=&\sigma \ob_{A^\prime}-
\kappa\ib_{A^\prime} \nonumber\\
\R_A&=&\rho\o_A-\kappa\i_A \nonumber\\
\T_{AA^\prime}&=& \tau \o_A\ob_{A^\prime}- \rho
\o_A\ib_{A^\prime}- \sigma \i_A\ob_{A^\prime}+
\kappa\i_A\ib_{A^\prime}
\end{eqnarray}
Under a transformation of the spin frame given by
\begin{eqnarray}
\o^A\gt\la\o^A \qquad \i^A\gt\la^{-1}\i^A+\bar a \o^A
\end{eqnarray}
these transform as
\begin{eqnarray}
\K&\gt& \la^3\lab \K \nonumber\\
\S_{A'}&\gt& \la^3\S_{A'} \nonumber\\
\R_A&\gt& \la^2\lab \R_A \nonumber\\
\T_{AA'}&\gt& \la^2 \T_{AA'}
\end{eqnarray}
They are therefore invariant under null rotations and have weight $\{\bp, \bq\}$
under spin and boost transformations given by
\begin{eqnarray}
\K&:\quad\{\3, \1\} \nonumber\\
\S&:\quad\{\3, \0\} \nonumber\\
\R&:\quad\{\2, \1\} \nonumber\\
\T&:\quad\{\2, \0\}
\end{eqnarray}
The role of the differential operators $\thorn$, $\edth$, $\thornp$
and $\edthp$ is played by new differential operators $\nthorn$,
$\nedth$, $\nthornp$ and $\nedthp$ which act on properly weighted
symmetric spinors to produce symmetric spinors of different valence
and weight. These operators may all be defined in terms of an
auxiliary differential operator $\cd_{ABA'B'}$ which is defined by
\begin{eqnarray}
&&   \cd_{ABA'B'}\etad \nonumber\\ & & = \o_A\ob_{A'}\nabla_{BB'}\etad
 \nonumber\\ & & \quad  - (\p\ob_{A'}\nabla_{BB'}\o_A   \ \ +\q\o_A\nabla_{BB'}\ob_{A'})\etad
 \label{ddi}
\end{eqnarray}
where $\boeta$ has weight $\{\p,\q\}$.

We will need to know the result of
contracting $\nthornp\boeta$ with $\o$ and $\ob$. We can write
 equation (\ref{ddi}) in the form
\begin{eqnarray}
& &\!\!\cd_{ABA'B'}\etad \nonumber\\& &= (\thornp\etad)\o_A\o_B\ob_{A'}\ob_{B'} \nonumber\\
& &\ -(\edthp\etad)\o_A\o_B\ob_{A'}\ib_{B'}-(\edth\etad)\o_A\i_B\ob_{A'}
\ob_{B'}\nonumber\\
& &\  \ -(\thorn\etad)\o_A\i_B\ob_{A'}\ib_{B'}\nonumber\\
& & \  \  \  \ +(\bp\i_A\ob_{A'}\T_{BB'}
+\bq\o_A\ib_{B'}\bar{\T}_{B'B})\etad
\end{eqnarray}
where $\thornp$, $\edthp$, $\edth$ and $\thorn$ are the ordinary GHP
operators applied to spinors. The new operators are obtained by
contraction with $\o$ and $\ob$, and symmetrizing.
\begin{eqnarray}
(\nthorn\boeta)_{AC_1\dots C_NA'C'_1\dots C'_{N'}}
&=& \sym\o^B\ob^{B'}\cd_{ABA'B'}\etad  \\
(\nedth\boeta)_{AC_1\dots C_NA'B'C'_1\dots C'_{N'}}
&=& \sym\o^B\cd_{ABA'B'}\etad  \\
(\nedthp\boeta)_{ABC_1\dots C_NA'C'_1\dots C'_{N'}}
&=& \sym\ob^{B'}\cd_{ABA'B'}\etad  \\
(\nthornp\boeta)_{ABC_1\dots C_NA'B'C'_1\dots C'_{N'}}
&=& \sym\cd_{ABA'B'}\etad
\end{eqnarray}
where $\displaystyle\sym$ indicates symmetrization over all free primed and
unprimed indices.
\medskip\noindent
In the case of a scalar field this gives
\begin{eqnarray}
(\nthornp\eta)_{ABA'B'}&=&
(\thornp\eta)\o_A\o_B\ob_{A'}\ob_{B'}
-(\edthp\eta-q\bar\tau\eta)\o_A\o_B\ob_{(A'}\ib_{B')} \nonumber\\
& & \ -(\edth\eta-p\tau\eta)\o_{(A}\i_{B)}\ob_{A'}\ob_{B'}
+(\thorn\eta-p\rho\eta-q\bar\rho\eta)\o_{(A}\i_{B)}\ob_{(A'}\ib_{B')} \nonumber\\
& &\  \  +(p\kappa\i_A\i_B\ob_{(A'}\ib_{B')}+q\bar\kappa\o_{(A}\i_{B)}\ib_{A'}\ib_{B'}\nonumber \\
& &\  \  \ -p\sigma\i_A\i_B\ob_{A'}\ob_{B'}-q\bar\sigma\o_A\o_B\i_{A'}\i_{B'})\eta
\label{thornp}
\end{eqnarray}

Contracting (\ref{thornp}) with $\ob^{B'}$ gives
\begin{eqnarray}
(\nthornp\eta)_{ABA'B'}\ob^{B'} &=
& \half\{(\nedthp\eta)_{ABA'}-q(\bar\tau\o_A\o_B\ob_{A'}
-\bar\rho\o_{(A}\i_{B)}\o_{A'} \nonumber\\
& & \quad -\bar\sigma\o_A\o_B\ib_{A'}
+\bar\kappa\o_{(A}\i_{B)}\i_{A'})\eta\} \nonumber\\
& =&  \half\{(\nedthp\eta)_{ABA'}-q\bar{\T}_{A'(A}\o_{B)}\eta\}
\end{eqnarray}
or in the compacted notation
\begin{eqnarray}
(\nthornp\eta)\cdot\ob=\half\{(\nedthp\eta)-q\bar\T\eta\} \label{thornp.ob}
\end{eqnarray}
Similar calculations give
\begin{eqnarray}
(\nthornp\eta)\cdot\o=\half\{(\nedth\eta)-p\T\eta\} \label{thornp.o}
\end{eqnarray}
\begin{eqnarray}
(\nedthp\eta)\cdot\o=\half\{(\nthorn\eta)-p\R\eta\} \label{edthp.o}
\end{eqnarray}
\begin{eqnarray}
(\nedth\eta)\cdot\ob=\half\{(\nthorn\eta)-q\bar\R\eta\} \label{edth.ob}
\end{eqnarray}
\begin{eqnarray}
(\nthornp\eta)\cdot\o\cdot\ob=\quarter\{(\nthorn\eta)-p\R\eta-q\bar\R\eta\}
\label{thornp.oob}
\end{eqnarray}
For a spinor  $\boeta$ the above contractions become more
complicated. For example for a valence (1,0)-spinor $\boeta_A$ of weight
$\{\p,\q\}$ we get
\begin{eqnarray}
(\nthornp\boeta)\cdot\o=\third\{\nthornp(\boeta\cdot\o)+
(\nedthp\boeta)-(\p-{\bf 1})\T\boeta\}\label{scont}
\end{eqnarray}

\medskip\noindent

Although the definition of the differential operators is quite
complicated, the fact that they take symmetric spinors to
symmetric spinors means that one can write down the equations in
an index free notation.

The Ricci equations, Bianchi equations and the commutators for the
general case are given in [2]. This complete system of equations
is completely equivalent to Einstein's equations, and to find
solutions to Einstein's equations this  system will therefore have
to be completely integrated. However, in view of the more
complicated nature of the operators in this formalism, some of the
information which resided in the Ricci equations in NP and/or GHP
formalisms is contained within the commutators in this formalism;
in particular these commutators contain inhomogeneous terms
explicitly dependent on the weight and valence of the spinor on
which they act. To extract all the information in the commutators
we need to apply them to, [15]

(i)\ \ four functionally
independent $\{0,0\}$ weighted real scalars,

(ii)\  one $\{p,q\}$ weighted complex scalar where $p\ne \pm q$,

(iii) one   valence $(1,0)$ spinor, $\I_A$ of  weight $\{{\bf -1},{\bf 0}\}$.

Of course, we can extract all the information by applying the commutators to
different (but essentially equivalent)  combinations of these scalars and
spinor; however the particular choices above are best suited to our integration
procedure since the four $\{0,0\}$ weighted real scalars will become the
coordinates, the complex scalar is given by the gauge field $\bar PQ$,
while the spinor $\I_A$ will be identified with the second dyad
spinor $\i_A$.

\section{The equations}

We restrict attention to the Petrov type N pure radiation spaces within the Kundt class of spacetimes (with a non-expanding and non-twisting null congruence); this means that when we choose $\o_A$ to be aligned with the propagation direction of the radiation
that the only remaining parts of the Riemann tensor are $\Phi_{22}$ and $\Psi_4$  which we simplify to $\Phi$ and $\Psi$ respectively, remembering that $\Phi$ is real; also  $\rho=0=\sigma=\kappa$.
In GIF, the Ricci spinor takes the form
\begin{eqnarray}
\bPhi_{ABA'B'}=\Phi\o_A\o_B\ob_{A'}\ob_{B'}
\end{eqnarray}
where $\Phi$ is a real scalar field of weight $\{-2,-2\}$.
and the Weyl spinor takes the form
\begin{eqnarray}
\bPsi_{ABCD}=\Psi\o_A\o_B\o_{C}\o_{D}
\end{eqnarray}
where $\Psi$ is a complex scalar field of weight $\{-4,0\}$. The
spin coefficients  in GIF satisfy
\begin{eqnarray}
\K &=0  \label{K} \nonumber\\
\S &=0  \label{S} \nonumber\\
\R & =0  \label{R}
\end{eqnarray}
but
\begin{eqnarray}
\T_{AA'}=\tau\o_A\ob_{A'}
\end{eqnarray}
Notice that $\tau , \Psi_4$ and $
\Phi_{22^\prime}$ are all invariant under the group of null rotations so
that they can be used instead of their GIF spinor equivalents; this gives a considerable simplification in the GIF notation.

\bigskip

The GIF equations are:

\bigskip

(i) GIF Ricci equations:

\bigskip
\begin{eqnarray}
\label{art1}
\bthorn \tau &=& 0\\
\bedth \tau &=& \tau^2\\
\bedthp \tau&=& \tau\overline{\tau}
\end{eqnarray}

\bigskip

(ii) GIF Bianchi identities:

\bigskip

\begin{eqnarray}
\label{art2}
\bthorn\Psi &=& 0\\
\bthorn\Phi &=& 0\\
\bedth\Psi -\bedthp\Phi &=& \tau \Psi -\overline{\tau}\Phi
\end{eqnarray}

\bigskip

(iii) GIF commutators (applied to an invariant spinor {\boldmath$\eta$}):

\bigskip

\begin{eqnarray}
\label{art3} (\bthorn \bthorn^\prime - \bthorn^\prime\bthorn)\mbox{\boldmath$\eta$}
&=& (\overline{\tau}\bedth + \tau\bedth^\prime)\mbox{\boldmath$\eta$}
\\
(\bthorn\bedth - \bedth\bthorn)\mbox{\boldmath$\eta$}&=& 0
\\
(\bedth\bedth^\prime - \bedth^\prime\bedth)\mbox{\boldmath$\eta$} &=& 0
\\
(\bthorn^\prime\bedth - \bedth\bthorn^\prime)\mbox{\boldmath$\eta$} &=&
-\tau\bthorn^\prime\mbox{\boldmath$\eta$} -\Phi(\mbox{\boldmath$\eta$}\cdot o)-
\overline{\Psi}(\mbox{\boldmath$\eta$}\cdot \overline{o})
\end{eqnarray}

\bigskip

These GIF equations  contain all the information for  the Type N pure
radiation metrics. We emphasise that we assume throughout that $\tau\ne 0 $.

\section{The integration procedure}

\subsection{Preliminary rearrangement.}

The spin coefficient $\tau$ will supply one real zero-weighted scalar $(\tau \bar \tau)$ and one real $\{1,-1\}$-weighted scalar $(\tau / \bar \tau)$.
However to keep the presentation of subsequent calculations as simple as possible, it will be convenient to rearrange slightly, and following \cite{edgar}, use instead the zero-weighted real scalar,
\begin{equation}
A=\frac{1}{\sqrt{2\tau\overline{\tau}}}.\label{art22}
\end{equation}

and the complex scalar $P$,
\begin{equation}
P=\sqrt{\frac{\tau}{2\overline{\tau}}
},\label{art20}
\end{equation}
 where $P$ is a scalar of weight $\{1,-1\}$, and $P\bar P ={1\over 2}$.

We are assuming $\tau\ne 0$, and so $A$ and $P$ will always be defined and different from zero.

These choices give the two simple partial tables
\begin{eqnarray}
\label{artP}
\bthorn P &=& 0\nonumber \\
\bedth P &=& 0\nonumber \\
\bedthp P &=& 0
\end{eqnarray}
and
\begin{eqnarray}
\label{artA}
\bthorn A &=& 0\nonumber \\
\bedth A &=&-P\nonumber \\
\bedthp A &=& -\overline{P}
\end{eqnarray}

We could now complete these tables with the fourth operator $\bthornp$ by introducing  unknown  spinors   and then applying each of the commutators to the completed tables of $P$ and $A$ respectively; unfortunately, the calculations soon become quite complicated.

Alternatively we could look for the other part of the weighted complex scalar by using expressions such as the $\{-1,-1\}$-weighted  scalar ${\Phi\over \tau \bar \tau}$ used in \cite{edgar}.  Again, calculations soon get quite complicated.

\subsection{Finding tables for the complex scalar $\bar PQ$, and the spinor ${\bf I}$  and applying commutators.}

So instead we examine the commutators and check whether their comparatively simple  structure suggests the existence of some simple table(s). In particular we know that we require a table for a real  weighted ($p= q \ne 0 $)  scalar to combine with $P$ to give a non-trivial weighted scalar  of weight $p\ne \pm q$. From the structure of the commutators we are motivated to consider the  simplest possible partial table for  a $\{-1,-1\}$-weighted  scalar $Q$ annihilated by the first three operators
\begin{eqnarray}
\label{art29} \bthorn Q &=& 0\nonumber\\
\bedth Q&=&0\nonumber\\
\bedthp Q &=&0
\end{eqnarray}
which leads us to consider the  table for $\bar PQ$
\begin{eqnarray}
\label{art29PQ} \bthorn (\bar PQ) &=& 0\nonumber\\
\bedth (\bar PQ)&=&0\nonumber\\
\bedthp (\bar PQ) &=&0\nonumber\\
\bthornp ( \bar PQ)
&=&- {Q\over A}{\bf I} \end{eqnarray}

where we have completed the table with  some spinor ${\bf I}$, which is as yet
undetermined.  (We have introduced  the weighted factor
${-Q\over A}$ in the above definition for ${\bf I}$ simply for convenience in
later calculations.)

It follows from (\ref{thornp.ob}) and (\ref{thornp.o}) that
\begin{eqnarray}
{\bf I} \cdot \ob  =-{ A\over Q} \bigl(\Ph'(\bar P Q)\bigr)\cdot
\ob
  =-{ A\over Q} \D'(\bar P Q)  =0 \label{I.ob}
\end{eqnarray}
\begin{eqnarray}
{{\bf I} \cdot  \o  =-{ A\over Q} \bigl(\Ph'(\bar P Q)\bigr)\cdot  \o
  =-{ A\over Q} \Bigl(\D(\bar P Q)+2\tau(\bar P Q)\Bigr)   = -1 }
\label{I.o}
\end{eqnarray}
Hence ${\bf I} $ is a $(1,0)$ valence spinor, and from
\begin{eqnarray}
\bigl(\Ph'(\bar P Q)\bigr)_{ABA'B'}  =-{Q\over A}{\bf I}_{(A}\o_{B)}
\ob_{A'}\ob_{B'}
\end{eqnarray}
we conclude that its weight is  $\{-\1,\0\}$.

So now we have to apply the commutators to the table for $ (\bar P Q)$ which
yields a partial table   for the spinor ${\bf I}$; the complete table can be
written as
\begin{eqnarray}
\label{art29I} \bthorn \bf {I} &=& 0\nonumber\\
\bedth \bf {I}&=&0\nonumber\\
\bedthp \bf {I} &=&0\nonumber\\
\bthornp  \bf {I}
&=&{\bar P Q^2\over A} {\bf W}
\end{eqnarray}
where the spinor ${\bf W}$ is as yet undetermined. (The factor ${\bar P Q^2\over A}$ is again just to improve efficiency of presentation.)
It follows from (\ref{scont}) that
\begin{eqnarray}
{\bf W} \cdot \ob  ={ A\over \bar P Q^2} \bigl(\Ph'{\bf
I}\bigr)\cdot
\ob
 = 0
\label{W.ob}
\end{eqnarray}
\begin{eqnarray}
{\bf W} \cdot  \o  =-{ A\over \bar P Q^2} \bigl(\Ph'{\bf
I}\bigr)\cdot
 \o
  = {1\over Q^2 \bar P^2} {\bf I}
\label{W.o}
\end{eqnarray}
Hence
\begin{eqnarray}
{\bf W }= -{1\over 2 \bar P^2Q^2} {\bf I}^2 + W \label{W}
\end{eqnarray}
where ${\bf W} $ is a $(2,0)$ valence spinor of weight   $\{\2,\0\}$, and $W$ is a
zero-weighted complex scalar.

We next have to apply the commutators to the previous table for ${\bf I}$ after the substitution (\ref{W}); we obtain a partial table for the
zero-weighted complex scalar $W$,
\begin{eqnarray}
\label{art29W}
\bthorn { W}&=&0\nonumber\\
\bedth { W}&=&-\frac{A}{\overline{P}Q^2}\Phi\nonumber\\
\bedthp { W}&=&-\frac{A}{\overline{P}Q^2}\Psi
\end{eqnarray}

When the commutators are applied to this partial table we obtain
\begin{eqnarray}
%\label{art2}
\bthorn\Psi &=& 0\\
\bthorn\Phi &=& 0\\
\bedth\Psi -\bedthp\Phi &=& \tau \Psi -\overline{\tau}\Phi
\end{eqnarray}
which are precisely the three Bianchi equations (\ref{art2})-(28).
Hence our guess for a table for $Q$ is completely compatible with
the commutators and all the other  equations. Furthermore,
checking the compatibility of our choice of table for
$\overline{P}Q$ led to a table for the spinor ${\bf I}$, whose
compatability we have also checked via the commutators.

So we have obtained two of the core elements required in our
analysis --- a weighted scalar $\overline{P}Q$ and a new spinor
${\bf I}$ which is not parallel to $\o$ --- and used them to
extract information from the commutators.

\subsection{Finding tables for two coordinate candidates and  applying commutators to each.}

We can now use $Q$ and ${\bf I}$ as defined above to complete the  table for  $A$, respecting spinor valences and weights and identities  (\ref{thornp.ob}) and (\ref{thornp.o}), and obtain
\begin{eqnarray}
\label{tableA}
\bthorn A &=& 0\nonumber \\
\bedth A &=&-P\nonumber \\
\bedthp A &=& -\overline{P}\nonumber\\
\bthornp A &=& P{\bf I} +\overline{P}\overline{\bf I}+\frac{Q}{A}N,
\end{eqnarray}
where $N$ is a real zero-weighted scalar, which is as yet undetermined.

 The
application of the  commutators to this table determines a table for the
scalar $N$,
\begin{eqnarray}
\label{art29a}
\bthorn N&=&-\frac{1}{Q}\nonumber\\
\bedth N&=&\frac{1}{Q}\overline{{\bf I}}\nonumber\\
\bedthp N&=& \frac{1}{Q}{\bf I}\\
\bthornp N&=&-\frac{1}{Q}{\bf I} \overline{{\bf I}} +{Q L\over A}.\nonumber
\end{eqnarray}
where $L$ is a real zero-weighted scalar, which is as yet undetermined.

The
application of the commutators to this table determines a partial table for the
scalar $L$,
\begin{eqnarray}
\label{art26c}
\bthorn L&=&0\nonumber\\
\bedth L&=&P\overline{W}\nonumber\\
\bedthp L&=&\overline{P}W
\end{eqnarray}
\medskip
At this stage we will review what we have obtained so far. We have tables for ${\bf I}, P, Q, A, N$ respectively, to all of which we have applied the commutators; this has yielded the further partial tables for $L$ and for (complex) W.  Since $A$ and $N$ are real zero-weighted scalars to which we have applied commutators they are two obvious candidates as coordinates.  (Of course it is necessary to confirm that these scalars are functionally independent, a fact which would seem obvious from the structure of the right hand side of their respective tables; however, we will take care to confirm this explicitly when we translate these tables into the purely scalar formalism.)

 If we wish to adopt either of the three remaining real zero-weighted scalars ($L $ or the real or the imaginary part of $W$) as coordinates, it would be necessary to complete the two tables and then apply the commutators to each; calculations then begin to get complicated.

 \subsection{Finding tables for two more coordinate candidates and applying commutators.}

Alternatively, we can go back to the commutators to see if they suggest  simple tables for the remaining  coordinates

We begin with the very simple table for a zero-weighted scalar $T$,
\begin{eqnarray}
\label{artB}
\bthorn T&=&0\nonumber\\
\bedth T&=&0\nonumber\\
\bedthp T&=&0\nonumber\\
\bthornp T&=&{Q\over A}
\end{eqnarray}
and we  easily confirm that this table  is  internally consistent,
as well as compatible with  the commutators and with all the other
equations.

Next we try the slightly more complicated
 table for  a real zero-weighted scalar $B$,
\begin{eqnarray}
\label{artB1}
\bthorn B&=&0\nonumber\\
\bedth B&=&-iP\nonumber\\
\bedthp B&=&i\overline{P}\\
\bthornp B&=&i(PI-\overline{P}\overline{I})\nonumber
\end{eqnarray}
and again confirm that this table is internally consistent, as
well as compatible with  the commutators and with all the other
equations.

\medskip

So we now have four coordinate candidates $A, N, T, B$, but it is necessary to confirm that these are functionally independent before we can adopt them formally as coordinates. Since these four scalars  are zero-weighted, we can easily rewrite their tables  using the   NP operators. We note that there are also the tables for $P$, $Q$ and ${\bf I}$, but, since these contain only gauge information, these tables play no role in the construction of the metric, and we do not consider them further.

\subsection{The tables in terms of scalar operators.}

If we identify the spinor ${\bf I}$ with the second dyad spinor $\i$, then the four tables for the zero-weighted coordinate candidates $T,A,N,B$ can be easily translated into the familiar NP operators,
\begin{eqnarray}
\label{art38np}
DT&=&0\nonumber\\
\delta T&=&0\nonumber\\
\delta^\prime T&=&0 \\
\Delta T&=&\frac{Q}{A} .\nonumber
\end{eqnarray}
\begin{eqnarray}
\label{art26np}
DA&=&0\nonumber\\
\delta A&=&-P\nonumber\\
\delta^\prime A&=&-\overline{P}\\
\Delta A&=&\frac{Q}{A}N\nonumber
\end{eqnarray}
\begin{eqnarray}
\label{art29anp}
DN&=&-\frac{1}{Q}\nonumber\\
\delta N&=&0 \nonumber\\
\delta^\prime N&=&0\\
\Delta N&=&\frac{Q}{A} L\nonumber
\end{eqnarray}
\begin{eqnarray}
\label{artBnp}
DB&=&0\nonumber\\
\delta B&=&-iP\nonumber\\
\delta^\prime B&=&i\overline{P}\\
\Delta B&=&0\nonumber
\end{eqnarray}

 A simple
observation of the determinant formed from the four vectors taken from  the right hand sides of each of the four tables
 respectively, confirms that  all four scalars are
functionally independent; so we formally adopt them as coordinates.

We note that these four tables for $T,N,A,B$ are not strictly involutive in these scalars; there occur also the  real scalars $L$  and $W$ which satisfy respectively
\begin{eqnarray}
\label{art26cnp}
DL&=&0\nonumber\\
\delta L&=&P\overline{W}\nonumber\\
\delta^\prime L&=&\overline{P}W
\end{eqnarray}
\begin{eqnarray}
\label{art29d}
D W&=&0\nonumber\\
\delta W&=&-\frac{A}{\overline{P}Q^2}\Phi\nonumber\\
\delta^\prime W&=&-\frac{A}{\overline{P}Q^2}\Psi
\end{eqnarray}

\subsection{Using coordinate candidates as coordinates.}

If we now make the choice of the coordinate candidates $T,N,A,B$ as coordinates
\begin{eqnarray} t=T,  \qquad n=N, \qquad a=A, \qquad b=B\nonumber
\end{eqnarray}
the tetrad vectors can be obtained in the $t,n,a,b$ coordinates
from their respective tables as follows:
\begin{eqnarray}
\label{tetrad}
l^i&=&(0,-\frac{1}{Q},0,0)\nonumber\\
n^i&=&\frac{Q}{a}(1,L,n,0)\nonumber\\
m^i&=&P(0,0,-1,-i)\\
\overline{m}^a&=&\overline{P}(0,0,-1,i)\nonumber
\end{eqnarray}
and we can write out the NP
derivatives as follows:
\begin{equation}
D=-\frac{1}{Q}\frac{\partial}{\partial n} \label{D}\nonumber
\end{equation}

\begin{equation}
\delta=-P(\frac{\partial}{\partial a}+i\frac{\partial}{\partial
b})=-P\frac{\partial}{\partial \xi} \label{delta}\nonumber
\end{equation}

\begin{equation}
\Delta=\frac{Q}{a}\frac{\partial}{\partial t}+L
\frac{\partial}{\partial n}+\frac{Qn}{a}\frac{\partial}{\partial
a}.\label{Delta}
\end{equation}
where it will be  convenient to write $\xi=a+ib$.

\bigskip

The metric $g$ can now be constructed using
\begin{equation}
g^{ij}=2l^{(i}n^{j)}-2m^{(i}\overline{m}^{j)}.\label{art40}
\end{equation}
to give in coordinates $t,n,a,b$,
\begin{equation}
g^{ij}= \left(
\begin{array}{lccr}
0 & -1/a & 0 & 0 \\
-1/a & L/a & -n/a & 0 \\
0 & -n/a & -1 & 0\\
0 & 0 & 0 & -1 \label{art41}
\end{array}
\right),
\end{equation}

From (\ref{art26cnp})
 we have $\bar W (t, \xi, \bar\xi)= L_{,\xi}$ where
 $L$ is now a function of the three coordinates $t, \xi, \bar\xi$,  and so from (60) we have
\begin{equation}
\Phi=\frac{Q^2}{2A}L_{,\overline{\xi}\xi} \label{partail1}
\end{equation}

\begin{equation}
\Psi=\frac{Q^2\overline{P}^2}{A}L_{,\overline{\xi}\overline{\xi}}
\label{partail2}
\end{equation}
\noindent

The above analysis is dependent on the condition that none of our  four  coordinate candidates could be  constant and still compatible with the equations; it is clear from considering the respective tables (\ref{art38np}), (\ref{art26np}), (\ref{art29anp}), (\ref{artBnp})  that this is the case. Therefore we can conclude that our analysis is complete in the sense that we have obtained the complete class of the metrics under consideration, and have not missed any 'special cases'.

The metric given by (\ref{art41}), is in very similar form to its
version  in \cite{ludwig}. We have made no restrictions on the
Riemann tensor components; therefore $\Psi=0$ gives the
conformally flat  special class, and $\Phi=0$ gives the vacuum
special class. Note however that the conformally flat class is in
different coordinates that in \cite{edgar}.

\section{Conclusion}

We have described here an alternative method within the GIF for obtaining
solutions of Einstein's equations and  applied it to a particular  class of Petrov type N  pure radiation spaces. The basic ideas of this method were developed in
previous work by Edgar and Vickers \cite{edgar}, where it was applied to
obtain the general conformally flat pure radiation metric.  That class of spaces were ideal because the absence of Killing vectors, in general, enabled a comparatively simple and direct analysis to be carried out; in the present paper such a direct analysis was not possible, and this is linked to a richer Killing vector structure.  So, in this application a modified approach was needed, and we have demonstrated how the introduction of {\it non-intrinsic} elements, imtroduced in compatability with the commutators, not only overcomes any problem created by the absence of sufficient coordinate candidates, but also can keep the calculations manageable. We anticipate that a major practical difficulty encountered when using this method resides in how to construct the second spinor ${\bf I}$ such that the subsequent tables will be as simple as possible; the procedure of introducing non-intrinsic elements gives us some control over this.

A big advantage of this procedure resides in the fact that  one has managed to avoid having
 to solve any differential equations. Another important advantage is that the complicated detailed
 calculations needed to keep track of coordinate transformations and gauge freedoms do not arise; this reduces
 the risk of direct computational mistakes, as well as the omission of special cases. It is an important development
 that
the GHP formalism now has computer support \cite{carm1},
\cite{carm2} and that attempts are being made to develop the GHP
integration procedure algorithmically in the programmes
\cite{carm3}; hopefully similar developments will occur soon for
GIF.

In \cite{edgar} it was easy (in the generic case) to draw
conclusions about Killing vector structure and invariant
classification; this was because only intrinsic elements of the
GIF were used. In this paper we have introduced a non-intrinsic
element early in our calculations, and this makes it more
difficult to draw conclusions about Killing vector structure and
the invariant classification. We shall discuss in detail how to
incorporate these consideration in general  into GIF in a
subsequent paper. At a more ambitious level,  we are also applying
this method to more  general vacuum type N metrics.

\end{document}